\documentclass[pra, twocolumn, showpacs, longbibliography]{revtex4-1}
\usepackage{epsfig}
\usepackage{amsmath}
\usepackage{dcolumn}
\usepackage{bm}
\usepackage{graphicx}
\usepackage{wrapfig}
\usepackage{graphicx}
\usepackage{epstopdf}
\usepackage{epsfig}
\usepackage{calrsfs}
\usepackage{IEEEtrantools}

\usepackage{amsmath}    
\usepackage{graphicx}   
\usepackage{verbatim}   
\usepackage{color}      
\usepackage{subfigure}  
\usepackage{hyperref}   

\newcommand{\nn}{\nonumber}

\newcommand{\ve}{\varepsilon}

\newcommand{\p}{\partial}

\newcommand{\Rbg}{Rabi-Bogoliubov}
\newcommand{\Bg}{Bogoliubov}
\newcommand{\BH}{Bose-Hubbard}

\newcommand{\wh}{\widehat}

\begin{document}



\title{Absence of Landau damping in driven three-component
Bose--Einstein condensate in optical lattices}
\author{Gavriil Shchedrin, Daniel Jaschke, and Lincoln D. Carr}
\affiliation{Colorado School of Mines, Golden, Colorado 80401, USA}

\begin{abstract}
We explore the quantum many-body physics of a three-component Bose-Einstein condensate (BEC) in an optical lattices driven by laser fields in $V$ and $\Lambda$  configurations. We obtain exact analytical expressions for the energy spectrum and  amplitudes of elementary excitations, and discover symmetries among them.  We demonstrate that the applied laser fields induce a gap in the otherwise gapless Bogoliubov spectrum.  We find that Landau damping of the collective modes above the energy of the gap is carried by laser-induced roton modes and is considerably suppressed compared to the phonon-mediated damping endemic to undriven scalar BECs.
\end{abstract}


\maketitle

Multicomponent Bose-Einstein condensates (BECs) are a unique form of matter that allow
one to explore coherent many-body phenomena in a macroscopic quantum system
by manipulating its internal degrees of freedom~\cite{kawaguchi2012spinor, ueda2010fundamentals, pethick2008bose}.
The ground state of alkali-based BECs, which includes $^{7}{\rm Li}$, $^{23}{\rm Na}$, and $^{87}{\rm Rb}$, is characterized by the hyperfine spin $F$, that can be best probed in optical lattices, which liberate its $2F+1$ internal components and thus provides a direct access to its internal structure~\cite{davis1995bose,  anderson1995observation, bradley1997bose,
chang2004observation, miesner1999observation, barrett2001all,
stenger1998spin, chang2005coherent}. Driven three-component $F=1$ BECs in $V$ and $\Lambda$
configurations (see Fig.~\ref{fig1}(b) and~\ref{fig1}(c))
are totally distinct from two-component BECs~\cite{pethick2008bose} due to the light interaction with three-level systems that results in the laser-induced coherence between excited states and ultimately leads to a number of fascinating physical phenomena, such as lasing without inversion (LWI)~\cite{PhysRevLett.62.2813, PhysRevLett.62.1033}, ultraslow light~\cite{hau1999light, PhysRevLett.82.5229}, and quantum memory~\cite{PhysRevLett.86.783}. The key technique behind these phenomena is electromagnetically induced transparency (EIT)~\cite{PhysRevLett.62.1033, PhysRevLett.66.2593}, which is based on the elimination of real and imaginary parts of the susceptibility upon applying a coherent resonant drive to a gas of three-level atoms, that opens a transparency window in otherwise optically opaque atomic media~\cite{harris1997today,RevModPhys.75.457, RevModPhys.77.633}.
The vanishing imaginary part of the susceptibility results in an extremely small group velocity of light, which led to the observation of unprecedented seven orders of magnitude slowdown of light propagation through a BEC of $^{23}$Na atoms~\cite{hau1999light}. The notion of non-dissipative dark-state polaritons not only yields a simple and elegant description of slow light phenomena~\cite{PhysRevLett.84.5094, PhysRevA.65.022314}, but also provides an efficient way to store and retrieve individual quantum states, i.e., quantum memory~\cite{PhysRevLett.86.783, julsgaard2004experimental, lvovsky2009optical,kielpinski2001decoherence}. Apart from physical phenomena achieved by the light-induced coherence in three-level systems, confined multicomponent BECs allowed the experimental realization of a number of fundamental physical concepts
including the observation of the spin Hall effect~\cite{li2014chiral}, creation of exotic magnetic~\cite{stamper2013spinor} and  topological states~\cite{choi2012observation, williams1999preparing}, and observation of Dirac monopoles~\cite{ray2014observation}. 

 \begin{figure}[t!]\label{fig1}
 \hspace{-5mm}
 \centering
\subfigure[]
{\includegraphics[width=0.4\columnwidth]{{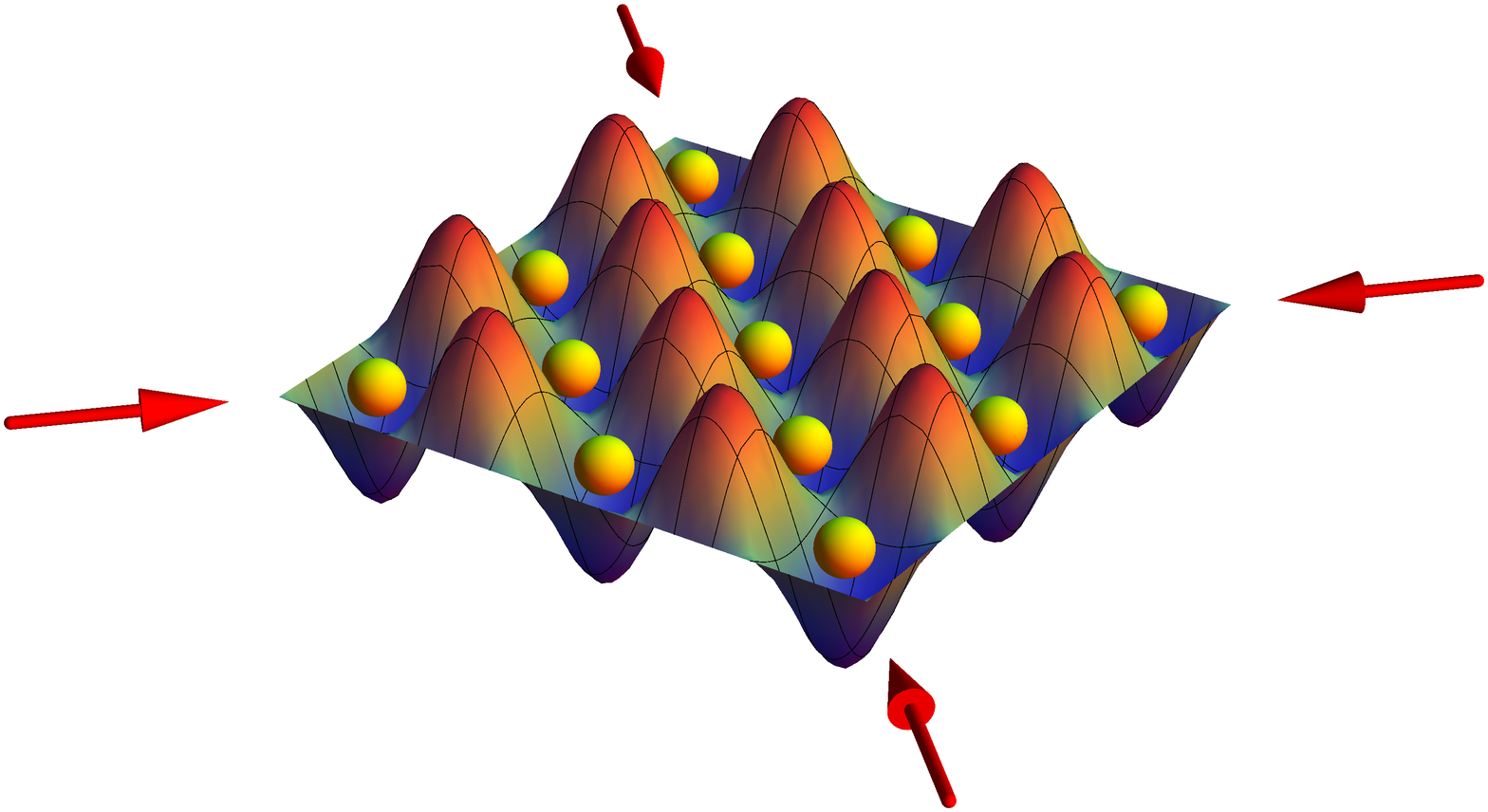}}}
\subfigure[]
{\includegraphics[width=0.32\columnwidth]{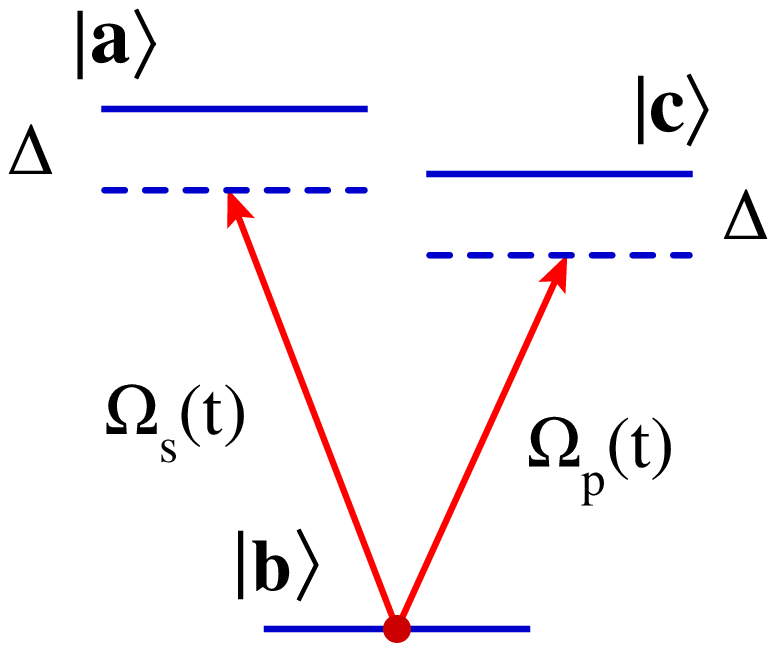}}
\subfigure[]
{\includegraphics[width=0.28\columnwidth]{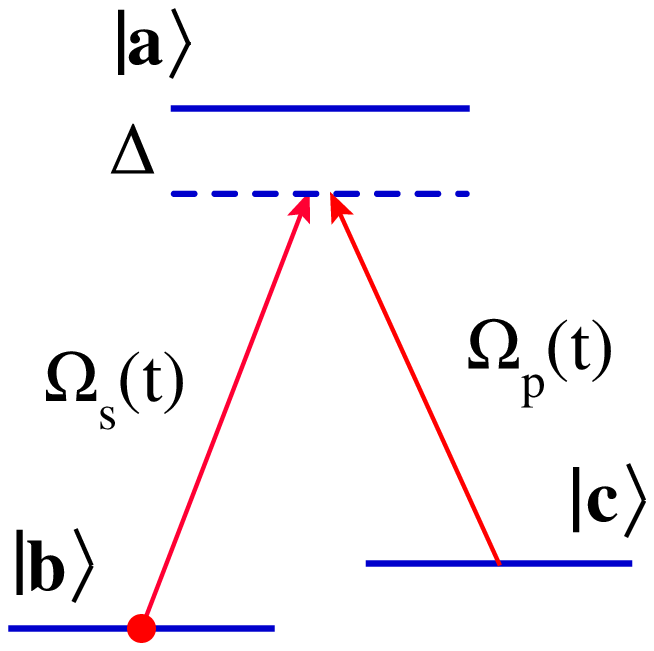}}
\caption{ {\it Three-Level Physics of BECs.} (a) Multicomponent $F=1$ BECs in optical lattices can have internal states $|F,m_F\rangle=\{|a\rangle,|b\rangle,|c\rangle\}$ in the well-known (b) $V$ and (c) $\Lambda$ configurations. A laser field $\Omega_{s}(t)$ drives the transition  between the ground state $|b\rangle$ and the excited state $|a\rangle$, while  $\Omega_{p}(t)$ induces the transition between states $|c\rangle$ and $|b\rangle$ for a $V$-system and $|c\rangle$ and $|a\rangle$ for the $\Lambda$ system. The laser fields are characterized by the Rabi frequencies $\Omega_{s}$ and $\Omega_{p}$ and equal detuning $\Delta$ from the excited states. Initially, the BEC is prepared in the ground state $|b\rangle$.}
\label{fig1}
\end{figure}

\begin{figure}[t!]
\centering
{\includegraphics[width=0.9\columnwidth]{{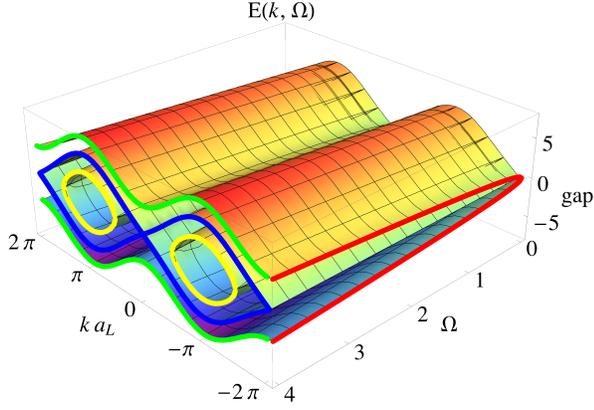}}}
\caption{
{\it Rabi-Bogoliubov (RB) Spectrum.} Energy gap for a 3-level BEC in $V$ and $\Lambda$ configurations as a function of driving frequency $\Omega=\sqrt{\Delta^{2}+\Omega_{s}^{2}+\Omega_{p}^{2}}$. $\Omega_{s}$ and $\Omega_{p}$ are the Rabi frequencies of the applied laser fields, and $\Delta$ is the
detuning (see Fig.{~\ref{fig1}}). The applied laser field induces a gap in the generalized {\Rbg} spectrum (red curve). The {\Rbg} spectrum is composed of three branches, separated by the standard gapless {\Bg} spectrum (blue curve) linear at small momenta $k$ (units of the lattice constant $a_{L}$). The first branch of the generalized {\Rbg} spectrum, which lies above the gap-less {\Bg} spectrum (green curve) is characterized by the energy gap. The second branch of the {\Rbg} spectrum (yellow curve) is characterized by the critical value of the momenta,
below which it does not have real-valued solution, and therefore, is bounded by the standard  {\Bg} spectrum.}
\label{fig2}
\end{figure}

However, access to these rich physical phenomena is limited or entirely excluded by the damping processes of the collective modes of the BEC~\cite{PhysRevA.88.063638, PhysRevA.88.031604, PhysRevLett.114.225303, phuc2013beliaev, sun2016ground}. The damping of the collective modes manifests itself in the metastable nature of spinor BEC, which dictates its properties and the many-body phenomena governed by it. Suppression of damping processes for collective excitations has been previously discovered experimentally and described theoretically in several BEC contexts, e.g., absence of Beliaev damping, which governs the decay process of a single collective mode into two collective excitations of a lower energy, for a quasi-2D dipolar gas~\cite{PhysRevA.88.031604,PhysRevA.88.063638, PhysRevA.90.043617}; the Quantum Zeno mechanism, responsible for a diminished decay rate of collective excitations in a quantum degenerate fermionic gas of polar molecules confined in optical lattices~\cite{yan2013observation}; and suppression of the Landau decay rate of collective excitations for a Bose-Fermi superfluid mixture~\cite{ferrier2014mixture, PhysRevLett.113.265304}. However, in all these systems
the energy spectrum is gapless for small momenta, and therefore, Landau damping is carried out predominantly by the phonons. In contrast to all these past studies, in this Letter we calculate exact analytical expressions for the Landau damping rate in spinor three-component BECs in optical lattices driven by microwave fields in both $V$ and $\Lambda$ configurations. The resulting generalized energy spectrum, Rabi-Bogoliubov (RB) amplitudes, and symmetries among them allow us to explore near-equilibrium BEC dynamics.  We find that the laser fields induce a gap (see Fig.~\ref{fig2}) in the energy spectrum, preventing collective excitation from Landau damping, and thus, enabling a metastable state in driven spinor BEC. The laser-induced gap in the energy spectrum results in zero group velocity and non-zero current for the collective excitations lying above the energy of the gap. Therefore, roton modes are induced, which significantly suppress Landau damping rate in spinor BECs compared to the phonon-mediated Landau damping in undriven scalar BECs.

We begin with the second-quantized Hamiltonian for a driven three-component BEC,
\begin{IEEEeqnarray}{l}\label{hamilt1}
H=\int{}{d\mathbf{r}}
\sum_{j=a,b,c}
\wh{\psi}_{j}^{\dagger}(\mathbf{r})
\left(
-\frac{\hbar^{2}}{2m}\nabla^{2}+
V(\mathbf{r})
-\mu_{j}
\right)
\wh{\psi}_{j}(\mathbf{r})
\nn\\
+
\frac{1}{2}
\int{}{d\mathbf{r}}
\sum_{j=a,b,c}
\wh{\psi}_{j}^{\dagger}(\mathbf{r})
\left(
\sum_{j'=a,b,c}
g_{jj'}
\wh{\psi}_{j'}^{\dagger}(\mathbf{r})
\wh{\psi}_{j'}(\mathbf{r})
\right)
\wh{\psi}_{j}(\mathbf{r})
\nn\\
+
\frac{\Omega_{s}}{2}
\int{}{d\mathbf{r}}
\left(
e^{i\Delta{t}}
\wh{\psi}_{a}^{\dagger}(\mathbf{r})
\wh{\psi}_{b}(\mathbf{r})
+
e^{-i\Delta{t}}
\wh{\psi}_{b}^{\dagger}(\mathbf{r})
\wh{\psi}_{a}(\mathbf{r})
\right)
\nn\\
+
\frac{\Omega_{p}}{2}
\int{}{d\mathbf{r}}
\left(
e^{i\Delta{t}}
\wh{\psi}_{c}^{\dagger}(\mathbf{r})
\wh{\psi}_{b}(\mathbf{r})
+
e^{-i\Delta{t}}
\wh{\psi}_{b}^{\dagger}(\mathbf{r})
\wh{\psi}_{c}(\mathbf{r})
\right)
\;\;\;
,
\end{IEEEeqnarray}
where we have chosen a $V$-configuration (see Fig.~\ref{fig1}b) for concreteness. Here
we introduced a Bose field operator $\wh{\psi}_{j}(\mathbf{r})$ which annihilates a particle
determined by the mass $m$, position $\mathbf{r}$, and internal state $j=b$ $(a,c)$ for a particle in the ground (excited) state. The lattice potential is assumed to have a simple cubic form, $V(\mathbf{r})=V_{0}\sum_{i=1}^{3}\sin^{2}(k_{L} r_{i})$, where $k_{L}=\pi/a_{L}$ is the lattice vector and $a_{L}$ is the lattice constant. The coupling constant $g_{jj'}$ determines the interaction between particles occupying the internal states $j$ and $j'$. 
The laser fields are characterized by the Rabi frequencies $\Omega_{s}$ and $\Omega_{p}$ and equal detuning $\Delta$ from excited states. Initially, the BEC is prepared in the ground state $|b\rangle$.

For sufficiently deep lattices, or alternatively in the long-wavelength approximation, one can safely adopt the lowest band approximation,
and perform expansion of the bosonic field operators $\wh{\psi}_{j}(\mathbf{r})$ in the Wannier basis $\wh{\psi}_{j}(\mathbf{r})=\sum_{n}\wh{ b }_{nj}w_{j}(\mathbf{r}-\mathbf{r}_{n})$. Throughout the paper, we will adopt the index convention, according to which the first argument of the field operator index denotes the site in an optical lattice, and the second argument indicates the internal state. Inserting this expansion into Eq.~(\ref{hamilt1}) we obtain,
\begin{IEEEeqnarray}{l}\label{hamilt2}
H=-\sum_{j=a,b,c}
\sum_{
\langle{m,n}\rangle}
J^{jj}_{mn}
\left(
\wh{ b }^{\dagger}_{mj}\wh{ b }_{nj}+
\wh{ b }^{\dagger}_{nj}\wh{ b }_{mj}
\right)
\\\nn
-
\sum_{j=a,b,c}
\mu_{j}
\sum_{n}
\wh{ b }^{\dagger}_{nj}\wh{ b }_{nj}
+
\sum_{j,j'=a,b,c}
\frac{U_{jj'}}{2}
\sum_{
n}
\wh{ b }^{\dagger}_{nj}
\wh{ b }^{\dagger}_{nj'}
\wh{ b }_{nj'}
\wh{ b }_{nj}
\\\nn
+
\frac{\Omega_{s}}{2}
\sum_{n}
\left(
e^{i\Delta{t}}
\wh{ b }^{\dagger}_{na}
\wh{ b }_{nb}
+
e^{-i\Delta{t}}
\wh{ b }^{\dagger}_{nb}
\wh{ b }_{na}
\right)
\\\nn
+
\frac{\Omega_{p}}{2}
\sum_{n}
\left(
e^{i\Delta{t}}
\wh{ b }^{\dagger}_{nc}
\wh{ b }_{nb}
+
e^{-i\Delta{t}}
\wh{ b }^{\dagger}_{nb}
\wh{ b }_{nc}
\right)
,
\end{IEEEeqnarray}
where we truncated the sum to the nearest neighbors, indicated by $\langle m,n\rangle$. Here the hopping integral is
\begin{equation}
J^{ij}_{mn}=
-\int
{d\mathbf{r}}
w^{*}_{i}(\mathbf{r}-\mathbf{r}_{m})
\left[
-\frac{\hbar^{2}}{2m}\nabla^{2}+
V(\mathbf{r})
\right]
w^{*}_{j}(\mathbf{r}-\mathbf{r}_{n})
,
\end{equation}
the on-site interaction is
\begin{IEEEeqnarray}{l}
U_{jj'}=
g_{jj'}
\int
{d\mathbf{r}}
w^{*}_{j}(\mathbf{r})
w^{*}_{j'}(\mathbf{r})
w_{j'}(\mathbf{r})
w_{j}(\mathbf{r})
,
\end{IEEEeqnarray}
 In order to formulate Eq.~(\ref{hamilt2}) in $k$-space we introduce the Fourier transform of the creation and annihilation operators,
\begin{IEEEeqnarray}{l}
\wh{ b }_{nj}=\frac{1}{\sqrt{N_{L}}}
\sum_{k}
\exp{[-i\mathbf{k}\mathbf{r}_{n}]}
\wh{a}_{kj}
,
\end{IEEEeqnarray}
where $N_{L}$ is number of lattice cites. In order to linearize the Fourier-transformed Hamiltonian given by Eq.~(\ref{hamilt2}) we expand the operators near their average values, i.e., $\wh{a}_{kj}=\langle{\wh{a}_{0j}}\rangle+(\wh{a}_{kj}-\langle{a_{0j}}\rangle)$. The average value of the field operator $\langle{a_{0j}}\rangle$ is given in terms of the number of 
particles occupying the zero momentum state $N_{0j}$, i.e. $\langle{\wh{a}_{0j}}\rangle=\sqrt{N_{0j}}$. The matrix, which describes coupling between particles identified by the internal state $j=\{a,b,c\}$  is given by,
\begin{equation}
\label{intmatrix1}
\left(
\begin{array}{ccc}
 n_{a}U_{aa} & \sqrt{n_{a}n_{b}}U_{ab} & \sqrt{n_{a}n_{c}}U_{ac} \\
 \sqrt{n_{b}n_{a}}U_{ba} & n_{b}U_{bb} & \sqrt{n_{b}n_{c}}U_{ac} \\
 \sqrt{n_{c}n_{a}}U_{ca} & \sqrt{n_{c}n_{b}}U_{cb} &  n_{c}U_{cc}
\end{array}
\right)\equiv{}
\left(
\begin{array}{ccc}
 u & s & t \\
 s & u & s \\
t & s & u
\end{array}
\right)
.
\end{equation}
Here we have introduced the average filling factor $n_{j}=N_{0j}/N_{L}$ for the particles characterized by the internal state $j$ and momentum $k=0$.  For brevity, we consider the simplified case, $s=0$ and $t=0$. However, the main physical results concerning the structure of the energy spectrum,
the amplitudes of elementary excitations of a driven three-component BEC, and symmetries among them, obtained in the most general case match the description here.
\begin{figure}[t!]\label{f1}
\includegraphics[angle=0, width=1.0\columnwidth]
{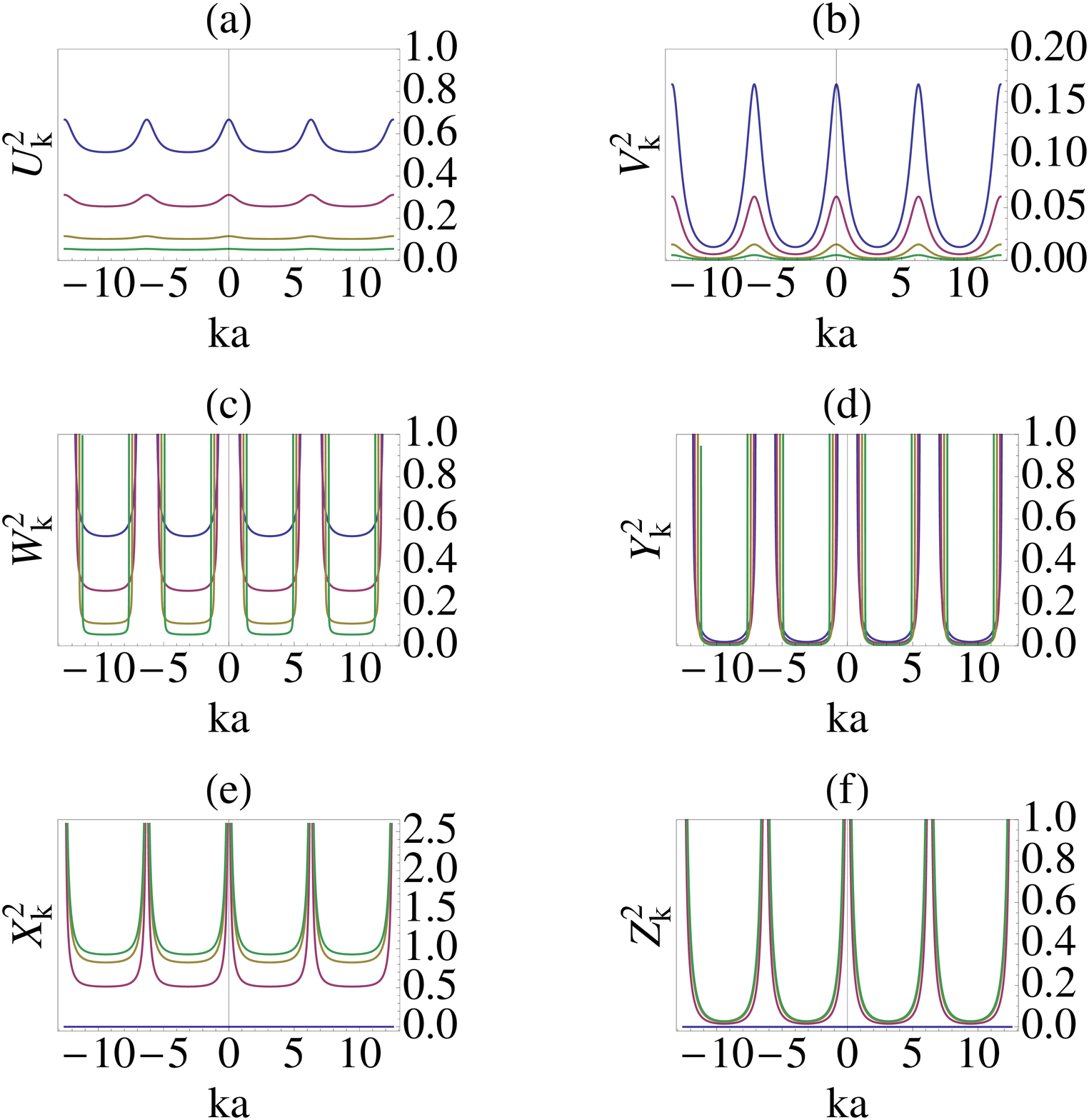}
 \caption{{\it Rabi-Bogoliubov Amplitudes as a Function of Rabi  Frequency.}
The standard {\Bg} amplitudes have poles at the location of the roots of {\Bg} spectrum. The laser field, which drives the BEC, creates a gap in the spectrum of elementary excitations. As a result, {\Rbg} amplitudes (a) $\mathcal{U}^{2}_{k}$ and (b) $\mathcal{V}_{k}^{2}$ are finite for all values of momenta $k$, given in the units of the lattice constant $a_{L}$. 
}
\label{ampiltudes1}
\end{figure}

The diagonalization of the Fourier-transformed {\BH} Hamiltonian can be accomplished via the generalized {\Bg} transformation. This transformation is carried out by the quasi-particle operators $\widehat{\alpha}_{k,a}$, $\widehat{\zeta}_{kc}$, and $\widehat{\beta}_{k,b}$,
that annihilate a particle occupying the internal state $j=a$, $j=c$ (excited states) and $j=b$ (ground state), correspondingly. The transformation from the particle to the quasi-particle basis is given by the linear combination, $\widehat{a}_{kj}=
\mathcal{U}_{k}\widehat{\alpha}_{ka}+
\mathcal{V}^{*}_{k}\widehat{\alpha}_{-ka}^{\dagger}+
\mathcal{W}_{k}\widehat{\beta}_{kb}+
\mathcal{Y}^{*}_{k}\widehat{\beta}_{-kb}^{\dagger}
+\mathcal{X}_{k}\widehat{\zeta}_{kc}+
\mathcal{Z}^{*}_{k}\widehat{\zeta}_{-kc}^{\dagger}
$.
The generalized {\Rbg} amplitudes 
(see Fig.~\ref{ampiltudes1}) are the subject to a constraint $\mathcal{U}_{k}^2-\mathcal{V}_{k}^2+\mathcal{W}^{2}_{k}- \mathcal{Y}^{2}_{k} +\mathcal{X}^{2}_{k} - \mathcal{Z}^{2}_{k}=1$, which ensures the bosonic commutation relation for the quasi-particle creation and annihilation operators, i.e.,
$[\wh{\alpha}_{kj}, \wh{\alpha}^{\dagger}_{-k,j'} ]=\delta_{jj'}$, and 
$[\wh{\beta}_{kj}, \wh{\beta}^{\dagger}_{-k,j'} ]=\delta_{jj'}$, and 
$[\wh{\zeta}_{kj}, \wh{\zeta}^{\dagger}_{-k,j'} ]=\delta_{jj'}$. 
In the basis of quasiparticle creation and annihilation operators, the {\BH} Hamiltonian acquires a diagonal form, 
\begin{IEEEeqnarray}{l}
H_{\text{eff}}=\\\nn
\frac{1}{2}\sum_{k}
\left(
E_{a}(k)
\widehat{\alpha}_{a,k}^{\dagger}
\widehat{\alpha}_{a,k}+
E_{b}(k)\widehat{\beta}_{b,k}^{\dagger}
\widehat{\beta}_{b,k}
+
E_{c}(k)\widehat{\zeta}_{b,k}^{\dagger}
\widehat{\zeta}_{b,k}
\right)
,
\end{IEEEeqnarray}
where $E_{a}(k)$, $E_{b}(k)$, and $E_{c}(k)$ are the three branches of the
{\Rbg} energy spectrum, see Fig.~\ref{fig2}, for both $V$ and $\Lambda$ configurations. In the case of the $V$-configuration, the energy spectrum
is obtained from the condition $\det{[M_{V}-\mathbf{1}(E/2)]}=0$, where 
%
\begin{IEEEeqnarray}{l}
M_{V}=\\\nn
\left(
\begin{array}{cccccc}
 h_{k}+\Delta  & -\frac{\Omega _s}{2} & 0 & u & 0 & 0 \\
 -\frac{\Omega _s}{2} & h_{k} & -\frac{\Omega _p}{2} & 0 & u & 0 \\
 0 & -\frac{\Omega _p}{2} & h_{k}+\Delta  & 0 & 0 & u \\
 -u & 0 & 0 & -h_{k}-\Delta  & \frac{\Omega _s}{2} & 0 \\
 0 & -u & 0 & \frac{\Omega _s}{2} & -h_{k} & \frac{\Omega _p}{2} \\
 0 & 0 & -u & 0 & \frac{\Omega _p}{2} & -h_{k}-\Delta
\end{array}
\right).
\end{IEEEeqnarray}
%
Here, the tunneling parameter $h_{k}=u + t_{k}$, with $t_{k} = 4J\sin^{2}\left({ka_{L}}/{2}\right)$, is given in terms of the tunneling amplitude $J\equiv{}J^{jj}_{mn}$, momentum $k$,
and the lattice constant $a_{L}$.

Thus, the {\Rbg} spectrum is 
%
\begin{align}\label{eigenvsys1}
E^{V}_{a,\pm}(k)&=\pm
{\sqrt{4\ve_{k}^{2}+4 \left(t_k+u\right) \Delta +
 \Delta ^2+\Omega^{2}+2\sigma}}
 ,
\nn\\
E^{V}_{b,\pm}(k)&=\pm
{\sqrt{4 \ve_{k}^{2}+4 \left(t_k+u\right) \Delta +
 \Delta ^2+\Omega^{2}-2\sigma}}
 ,\nn\\
 E^{V}_{c,\pm}(k)&=\pm\sqrt{\left(t_k+\Delta \right) \left(t_k+2 u+\Delta \right)}
\end{align}
where $\ve_{k}= \sqrt{t_k \left(t_k+2 u\right)}$ is the standard Bogoliubov spectrum~\cite{pethick2008bose}, $\Omega=\sqrt{\Delta^{2}+\Omega_{s}^{2}+\Omega_{p}^{2}}$ is the effective Rabi frequency of the combined laser fields $\Omega_{s}(t)$ and $\Omega_{p}(t)$
and $\sigma\equiv\left(2 \left(t_k+u\right)+\Delta \right)\Omega$.

We find a set of new symmetries that holds among the generalized {\Rbg} amplitudes, 
\begin{align}\label{symmetry1}
\mathcal{V}_{k}^2(E_{a},\Omega)
&=-
\mathcal{U}^{2}_{k}(-E_{a},\Omega)
,\\\nn
 \mathcal{W}^{2}_{k}(E_{b},\Omega)
 &=
 \mathcal{U}^{2}_{k}(E_{b},-\Omega)
 ,\\\nn
\mathcal{Y}^{2}_{k}(E_{b},\Omega)
&=-
\mathcal{U}^{2}_{k}(-E_{b},-\Omega)
,\\\nn
\mathcal{X}^{2}_{k}(E_{c})
&=-
\mathcal{Z}^{2}_{k}(-E_{c})
.
\end{align}
The new symmetries summarized by Eq.~(\ref{symmetry1}) are the direct generalization of the intrinsic symmetries of the standard {\Bg} amplitudes, i.e., $u^{2}_{k}(E)=-v^{2}_{k}(-E)$~\cite{pethick2008bose}. These symmetries generate the complete set of the generalized 
{\Rbg} amplitudes from a single amplitude $\mathcal{U}^{2}_{k}(E_{a},\Omega)$, which for a $V$-system is explicitly given by
%
\begin{equation}
\mathcal{U}^{2}_{k}(E_{a},\Omega) =
 \frac{\Omega _p^2+\Omega _s^2+\left(2 \left(t_k+u\right)+E^{V}_{a} \right)
 \left(\Omega-\Delta \right)}{4 E^{V}_{a} \Omega}
\end{equation}
%
%
and for the $\Lambda$-system,
\begin{equation}
\mathcal{U}^{2}_{k}(E_{a},\Omega) =
\frac{\Omega _s^2 \left[\Omega _p^2+\Omega _s^2+\left(2 \left(t_k+u\right)+E^{\Lambda}_{a} \right) \left(\Omega-\Delta \right)\right]}{4 E^{\Lambda}_{a} \left(\Omega _p^2+\Omega _s^2\right) \Omega}
\end{equation}
The symmetries in the $V$-system result in cancellation of the amplitudes $\mathcal{X}_{k}(E_{c})$ and $\mathcal{Z}_{k}(E_{c})$, while for the $\Lambda$-system we have
\begin{eqnarray}
\mathcal{X}^{2}_{k}(E_{c}) =
 \frac{ \Omega _p^2\left(t_k+u+E^{\Lambda}_{c}\right)}{2 E^{\Lambda}_{c} \left(\Omega _p^2+\Omega _s^2\right)}
\end{eqnarray}
Here $E^{\Lambda}_{a}(k)$, $E^{\Lambda}_{b}(k)$, and $E^{\Lambda}_{c}(k)$
are solutions of the eigenvalue problem for the $\Lambda$ configuration governed by
\begin{equation}
M_{\Lambda}=
\left(
\begin{array}{cccccc}
 h_{k}+\Delta  & -\frac{\Omega _s}{2} & -\frac{\Omega _p}{2} & u & 0 & 0 \\
 -\frac{\Omega _s}{2} & h_{k} & 0 & 0 & u & 0 \\
 -\frac{\Omega _p}{2} & 0 & h_{k} & 0 & 0 & u \\
 -u & 0 & 0 & -h_{k}-\Delta  & \frac{\Omega _s}{2} & \frac{\Omega _p}{2} \\
 0 & -u & 0 & \frac{\Omega _s}{2} & -h_{k} & 0 \\
 0 & 0 & -u & \frac{\Omega _p}{2} & 0 & -h_{k}
\end{array}
\right)
\end{equation}
The eigenvalues in $\Lambda$-configuration are given in terms of the energy spectrum for the $V$-system, Eq.~(\ref{eigenvsys1}),
\begin{IEEEeqnarray}{l}
E^{\Lambda}_{a}(k)=E^{V}_{a}(k)\\\nn
E^{\Lambda}_{b}(k)=E^{V}_{b}(k)\\\nn
E^{\Lambda}_{c,\pm}(k)=\pm\sqrt{t_k \left(t_k+2 u\right)}
\end{IEEEeqnarray}

In the long wavelength limit the {\Rbg} amplitudes $\mathcal{W}_{k}$, $\mathcal{Y}_{k}$ are purely imaginary. Therefore, we are left with the real-valued {\Rbg} amplitudes $\mathcal{U}_{k}(E)$, $\mathcal{V}_{k}(E)$, $\mathcal{X}_{k}(E)$, and $\mathcal{Z}_{k}(E)$,
that could be further simplified in case of resonant driving fields, i.e. $\Delta=0$. For the laser fields in the $V$-configuration we have,
\begin{IEEEeqnarray}{l}
\mathcal{U}^{V}_{k}(E_{a})=\sqrt{
[E_{a} +2 (t_k+u + \Omega_{0}/2)]/
(4 E_{a} )}
,
\\\nn
\mathcal{V}^{V}_{k}(E_{a})=-\sqrt{
[-E_{a} +2 (t_k+u+\Omega_{0}/2)]/
(4 E_{a})},
\end{IEEEeqnarray}
while for the $\Lambda$-system the {\Rbg} amplitudes can be simplified into,
\begin{IEEEeqnarray}{l}
\mathcal{U}^{\Lambda}_{k}(E_{a})=
\frac{\Omega_{s}}{\Omega_{0}}
\sqrt{
\frac{E_{a} +2 (t_k+u + \Omega_{0}/2)}
{4 E_{a} }}
,
\\\nn
\mathcal{V}^{\Lambda}_{k}(E_{a})=-
\frac{\Omega_{s}}{\Omega_{0}}
\sqrt{
\frac{-E_{a} +2 (t_k+u+\Omega_{0}/2)}
{4 E_{a} }}.
\end{IEEEeqnarray}
The amplitudes $\mathcal{X}^{\Lambda}_{k}(E_{c})=
{\Omega_{p}}/{\Omega_{0}}
u_{k}(E_{c})$ and $\mathcal{X}^{\Lambda}_{k}(E_{c})=
{\Omega_{p}}/{\Omega_{0}}v_{k}(E_{c})$ are given in terms of the standard {\Bg} amplitudes $u_{k}$ and $v_{k}$. The effective Rabi frequency simplifies into
$\Omega_{0}\equiv{}\Omega(\Delta=0)=\sqrt{\Omega_{s}^{2}+\Omega_{p}^{2}}$ and $E_{a}=\sqrt{\left(2 t_k+\Omega\right) \left(2 t_k+4 u+\Omega\right)}$ and $E^{\Lambda}_{c}=\sqrt{t_k \left(t_k+2 u\right)}$.

Introducing $E=E_{a}/2$, we obtain the following
expression for the Landau damping rate for the laser fields in the $V$ configuration,
\begin{align}\label{landauint1}
\Gamma^{V}_{L}&=
-\pi \hbar \omega_{q}
\frac{2\pi}{(2\pi\hbar)^{3}}
\left(
4 \sqrt{N}\frac{g_{jj}}{2}
\frac{\sqrt{\omega_{q} }}{\sqrt{2} \sqrt{u+s}}
\right)^{2}
\frac{1}{q}
\nn\\
&\times
 \beta
\frac{\p}{\p \beta}
\int{}
{dp}\;
\frac{1}{v_{g}}
\frac{p^{2}}{E}
\frac{1}{(e^{\beta  E}-1)}
\left(
\frac{3}{4}\frac{E}{(u+s)}
\right)^{2}
.
\end{align}
Here $\omega_{q}$ and $q$ are the frequency and momentum of the collective mode, respectively. In the $\Lambda$-configuration the Landau damping rate is
\begin{eqnarray}
\Gamma^{\Lambda}_{L}=(\Omega_{s}/\Omega_{0})^{2}\Gamma^{V}_{L} +
(\Omega_{p}/\Omega_{0})^{2}\Gamma_{L},
\end{eqnarray}
where $\Gamma_{L}$ is the usual Landau damping rate in the laser-free case~\cite{pitaevskii1997landau}. The Landau damping acquires a particularly simple form if we introduce the density of the laser-induced roton modes, 
\begin{align}\label{densityrot1}
\rho_{r}&=
\frac{4\pi}{3(2\pi\hbar)^{3}}
\int{}
{dp}\;
p^{2}
\frac{E^{2}}{v_{g}}
\left(-\frac{\p}{\p E}
\frac{1}{(e^{\beta  E}-1)}
\right)
\nn\\
&=
\frac{4\pi}{3(2\pi\hbar)^{3}}
\left(-\beta\frac{\p}{\p \beta}\right)
\int_{E_{0}}^{\infty}
{dE}\;
\frac{p^{2}}{v_{g}^{2}}
\frac{E}{(e^{\beta  E}-1)}
.
\end{align}
Here we have introduced the group velocity $v_{g}={\p E(p)}/{\p p}$, the Boltzmann factor 
$\beta=1/k_{B}T$, and the Boltzmann constant $k_{B}$. The Taylor expansion of the energy  around zero momentum returns, $E\simeq{}E_{0}+E_{2}p^{2}/2$,  where that gap in the spectrum is $E_{0}=\sqrt{\Omega \left(4 u+\Omega \right)}/2$, and the curvature of the spectrum is $E_{2}=\left(2 u+\Omega \right)/
[m^{*} \sqrt{\Omega \left(4 u+\Omega\right)}]$. The effective mass is given by $m^{*}={1}/(Ja_{L}^{2})$. 

Finally, we can express the rate of Landau damping for a three-component BEC driven by the laser fields in a $V$-configuration in terms of the density of the laser-induced roton modes,
\begin{equation}\label{landaurot1}
\Gamma^{V}_{L}=
\theta(\hbar\omega_{q}-E_{0})
\frac{27\pi}{16}
\hbar\omega_{q}
\frac{\rho_{r}}{\rho(\omega_{q})}
.
\end{equation}
Here the spectral density of the collective modes, $
\rho(\omega_{q})=
q(u+s)^{3}/(g_{jj}^{2}N\omega_{q})$, is given in terms the $q=
\sqrt{{2(\hbar\omega_{q}-E_{0})}/{E_{2}}}$. We immediately find that the collective modes characterized by the energy not exceeding the energy of the gap ($\hbar\omega_{q} < E_{0}$) are free from Landau damping, i.e., $\Gamma_{L} = 0$. Therefore, the gap in the energy spectrum produced by the applied laser fields effectively protects low-lying collective modes from Landau damping. For the collective modes lying above the energy of the gap, the Landau damping rate scales with the density of the laser-induced roton modes, which in the limit of low temperatures behaves as $\rho_{r}\simeq{}\beta^{-2}$. In the limiting case of laser-free condensate, i.e., $\Omega_{s}=\Omega_{p}=0$, the {\Rbg} spectrum simplifies to the standard {\Bg} spectrum. As a result, laser-modified Landau damping rate Eq.(\ref{landaurot1}) reduces to the well-known result \cite{pitaevskii1997landau} for the phonon-mediated Landau damping of the collective modes in scalar BEC,
$\Gamma_{L}(\Omega_{s} = \Omega_{p} = 0)=
\frac{27\pi}{16}
\hbar\omega_{q}
\frac{\rho_{n}}{\rho}\simeq{}\frac{1}{\beta^{4}}$,
defined in terms of the density of a phonon gas $\rho_{n}=2\pi^{2}T^{4}/({45\hbar^{3}c^{5}})$ \cite{landau1980statistical}.
Thus, Landau damping rate of the collective excitations in a driven three-level BEC is
significantly slowed down compared to scalar laser-free BEC, where damping processes are mediated by phonons.

Experimentally, the absence of Landau damping in driven three-component condensate can be verified by means of the two-photon Bragg spectroscopy. This technique was successfully applied in measuring Beliaev damping of the collective modes \cite{PhysRevLett.89.220401}, which revealed a complete absence of the collision of quasiparticles below a critical momentum in a BEC of $^{87}$ Rb atoms. In case of Beliaev damping of collective modes in a laser-free BEC, as well as in case of Landau damping in a laser-driven spinor condensate, both physical systems are characterized by a critical energy, below which collision of the collective modes and the corresponding damping processes are entirely excluded. Thus, we conclude that despite the fact that the collision of the quasiparticles reported in the experiment \cite{PhysRevLett.89.220401} was governed by Beliaev damping, we anticipate the same results for Landau damping of collective modes in a laser-driven three-component spinor Bose-Einstein condensates. 

In conclusion, we investigated the quantum many-body physics of a three-component BEC confined in optical lattices and driven by laser fields in both $V$ and $\Lambda$ configurations. We found that the applied laser fields create a gap in the spectrum that shields collective excitation of the condensate lying below the energy of the gap from Landau damping. Above the gap, Landau damping is proportional to the density of the laser-induced roton modes, and is substantially suppressed compared to the Landau damping rate in an undriven scalar condensate carried by the phonons.  This advance provides a prescription for the realization of electromagnetically induced transparency  and other exciting three-level phenomena in multicomponent Bose-Einstein condensates.

\begin{acknowledgments}
The  authors  gratefully  acknowledge  stimulating  discussions with Marc Valdez and Logan Hillberry. This material is based in part upon work supported by the US National Science Foundation under grant numbers PHY-1306638, PHY-1207881, and PHY-1520915, and the US Air Force Office of Scientific Research grant number FA9550-14-1-0287.
\end{acknowledgments}


%

\end{document}